# Tissue-Contrastive Semi-Masked Autoencoders for Segmentation Pretraining on Chest CT

Jie Zheng, Ru Wen, Haiqin Hu, Lina Wei, Kui Su, Wei Chen, Chen Liu, and Jun Wang

***Abstract*—Existing Masked Image Modeling (MIM) depends on a spatial patch-based masking-reconstruction strategy to perceive objects' features from unlabeled images, which may face two limitations when applied to chest CT: 1) inefficient feature learning due to complex anatomical details presented in CT images, and 2) suboptimal knowledge transfer owing to input disparity between upstream and downstream models. To address these issues, we propose a new MIM method named Tissue-Contrastive Semi-Masked Autoencoder (TCS-MAE) for modeling chest CT images. Our method has two novel designs: 1) a tissue-based masking-reconstruction strategy to capture more fine-grained anatomical features, and 2) a dual-AE architecture with contrastive learning between the masked and original image views to bridge the gap of the upstream and downstream models. To validate our method, we systematically investigate representative contrastive, generative, and hybrid self-supervised learning methods on top of tasks involving segmenting pneumonia, mediastinal tumors, and various organs. The results demonstrate that, compared to existing methods, our TCS-MAE more effectively learns tissue-aware representations, thereby significantly enhancing segmentation performance across all tasks.**

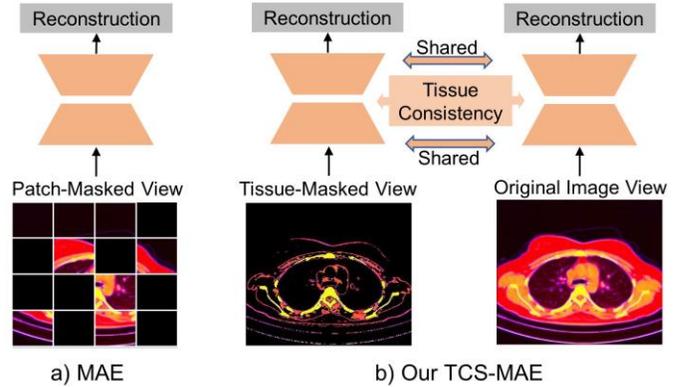

Fig. 1. (a) The MAE may fail to learn the tissue-aware representations through the patch-based masking way, and its over-masking procedure will destruct the input consistence between pretrain and downstream models. (b) Our proposed TCS-MAE is designed to learn the tissue-aware features by masking tissue regions instead of image patches. The original image view is incorporated to bridge the gap of the upstream and downstream models.

## I. INTRODUCTION

RELIABLE automatic segmentation of anatomical tissues or lesions in chest CT images is vital for the computer-aided diagnosis of many diseases such as pneumonia[1][2], lung cancers[3], and mediastinal tumors[4]. Deep learning methods like U-Net[5] and its variants[6]–[8] have significantly advanced medical image segmentation. However, training a robust segmentation model requires ample samples with pixel-wise annotations, which are often difficult to obtain due to their cost, time-consuming nature, and expertise required.

Recently, Masked Image Modeling (MIM) has become popular in the field of self-supervised learning (SSL), as it offers an effective means to learn visual representations from vast amounts of unlabeled images. The learned knowledge then can be transferred to downstream models to improve their performance. Existing MIM methods can roughly be divided into two categories: Pure masking methods[9][10] and hybrid methods[11]–[13]. The pure masking methods like MAE[9] steer the feature learning by randomly masking and then reconstructing image patches using an Autoencoder (AE). The AE model possesses an encoder-decoder architecture, making it naturally suitable for segmentation tasks. However, while these methods focus on spatial patches to learn discriminative features, they may fail to perceive holistic semantics.

Consequently, the hybrid methods were developed to empower the MAE with contrastive learning or other tricks in capturing more holistic features. For instance, Huang et al.[11] proposed the Contrastive Masked Autoencoder (CMAE) for stronger image modeling. They added a momentum encoder to the online MAE, which takes whole images as inputs to generate feature embeddings for contrastive learning against their online masked counterparts. This contrastive dual-encoder setup enables the capture of both holistic and local discriminative features, thus achieving better generalization performance.

Although the above methods have been proven effective for

This work was supported in part by National Natural Science Foundation of China under Grant 62101318 and part by Natural Science Foundation of Zhejiang Province of China under Grant LTGY24H180010. *Corresponding authors*: *K. Su, W. Chen, C. Liu, and J. Wang. Co-first authors*: *J. Zheng and R. Wen.*

J. Zheng, H. Q. Hu, N. L. Wei, K. Su, and J. Wang are with the Hangzhou City University, Hangzhou, China (e-mail: avengerjie@163.com; huhaiqin12@163.com; weiln@hzcu.edu.cn; suk@hzcu.edu.cn; wangjun@hzcu.edu.cn).

R. Wen, C. Liu are with the Department of Radiology, Southwest Hospital, Army Medical University (Third Military Medical University), Chongqing, 40038, China (e-mail: wenru@aifmri.com; liuchen@aifmri.com).

W. Chen is with the State key Lab of CAD&CG and the Laboratory of Art and Archaeology Image, Zhejiang University, China (e-mail: chenvis@zju.edu.cn)

modeling natural images, they may face limitations when applied for modeling chest CT images. On one hand, chest CT images normally contain large non-informative background regions that surround complicated anatomical tissues with diverse structures like vessels, bones, and muscles in the lung or mediastinal areas. These structures differ significantly in size and shape from objects in natural images. Additionally, CT images use Hounsfield Unit (HU) instead of grayscale values to represent tissues. The intensity of different tissues varies across a wide HU range, approximately from -1000 HU to 500 HU. Such complexities suggest that a spatial-patch-based masking strategy might be inefficient for capturing the semantic integrity of tissues (See Fig. 1a); On the other hand, MIM necessitates a large masking ratio (e.g., 75%) to attain effective representation capacity[9]. Since downstream models often receive original uncorrupted images, this over-masking operation can result in input discrepancies between the pretrained models and their downstream counterparts (referred to as upstream-downstream inconsistency issue), potentially hindering the transfer of knowledge to the downstream tasks.

To tackle the above challenges, we present a novel MIM method named Tissue-Contrastive Semi-Masked Autoencoder (TCS-MAE) tailored for segmentation pretraining on chest CT images. As illustrated in Fig. 1b, our TCS-MAE framework consists of two modules with shared weights: a tissue-based MAE and a conventional AE. The tissue-based MAE takes masked images as inputs, which are generated using an intensity-based masking strategy rather than the conventional patch-based masking. Specifically, this masking strategy segregates the HU spectrum into multiple intervals and masks them at random, leveraging the fact that different tissues are characterized by specific HU ranges. Since the tissue-based MAE tries to reconstruct the obscured tissue regions instead of patches during training, it is expected to capture more discriminative tissue representations. Moreover, the original AE is designed for two purposes: 1) to enhance the efficiency of tissue learning by enforcing consistency between the feature maps of the original image and its corresponding masked view; 2) to mitigate the upstream-downstream inconsistency issue by exposing the model to original, unaltered images throughout the training process.

Furthermore, as previously mentioned, the HU ranges for the lung and mediastinal regions differ in CT images, making it challenging to learn the representations of both simultaneously. To address this problem, we convert the chest CT images into two sets of grayscale images using a lung window and a mediastinal window. Then, the lung and mediastinal images along with their edge images are concatenated in the channel dimension to construct RGB images that serve as input to the TCS-MAE. This conversion facilitates the adaptation of existing SSL methods designed for natural images to the context of CT image representation learning.

To validate our method, we have conducted comprehensive experiments on three downstream tasks, including pneumonia segmentation, mediastinal tumor segmentation, and chest organ segmentation. The results demonstrate that, in comparison to existing pretraining methods, our TCS-MAE approach can learn tissue-aware representations more effectively and substantially improve the segmentation performance across all downstream tasks. Our primary contributions can be summarized as follows:

1) We propose a novel masked image modeling paradigm named TCS-MAE for self-supervised visual learning in CT images, which explores how to improve representations by integrating MAE with contrastive learning.

2) We present a new tissue-based masking strategy tailored for learning more holistic tissue representations in CT images compared to the conventional patch-based masking strategy.

3) We pretrained foundation models using more than 6000 normal CT scans, and extensive experiments demonstrated the effectiveness of our models based on various segmentation tasks. We will release these models to the community and contribute to the advancement of computer-aided chest CT image analysis.

## II. RELATED WORK

### A. Contrastive Learning

Contrastive Learning (CL) is one of the most popular SSL paradigms, which tries to extract consistent features by generating positive and negative image pairs. Chen et al.[14] proposed a Siamese framework named SimCLR for contrastive learning of visual representations that benefits from data augmentations. Caron et al.[15] further improved the Siamese representation learning using a novel similarity matching loss, and thus proposed a state-of-the-art (SOTA) SSL method called DINO. He et al.[16] formulated the contrastive learning as a dictionary look-up problem, and proposed Momentum Contrast (MoCo) for unsupervised visual representation learning.

Inspired by the above studies, some searchers have applied CL to medical image modeling. For example, Yan et al.[17] developed a CL-based SSL method named SAM for learning anatomical embeddings in radiological images. Liang et al.[18] proposed a contrastive cross-model pretraining strategy for small sample medical imaging. However, pure CL methods assume the downstream applications to be classifications and only train an encoder for feature learning. Therefore, they are not suitable for image segmentation tasks. To address this issue, Chaitanya et al.[19] designed a two-stage pretraining strategy for medical image segmentation: encoder-pretraining followed by decoder-pretraining. This strategy is effective but suffers from a relatively complex training pipeline.

### B. Generative Learning

Generative learning is another mainstream paradigm for modeling images, most of which rely on Auto-Encoder (AE) architectures. The basic goal of these methods is to recover images from their corrupted versions using an encoder-decoder structure. The corrupted images are obtained by adding disturbances such as noising[20], masking[21], rotation[22], or a mixed strategy[23]. The primary benefit of these methods lies in their straightforward pipelines for model training. However, as previously mentioned, the corrupting procedures might break the data intrinsic distributions and result in discrepancies between the input to the model during pretraining and downstream tasks.

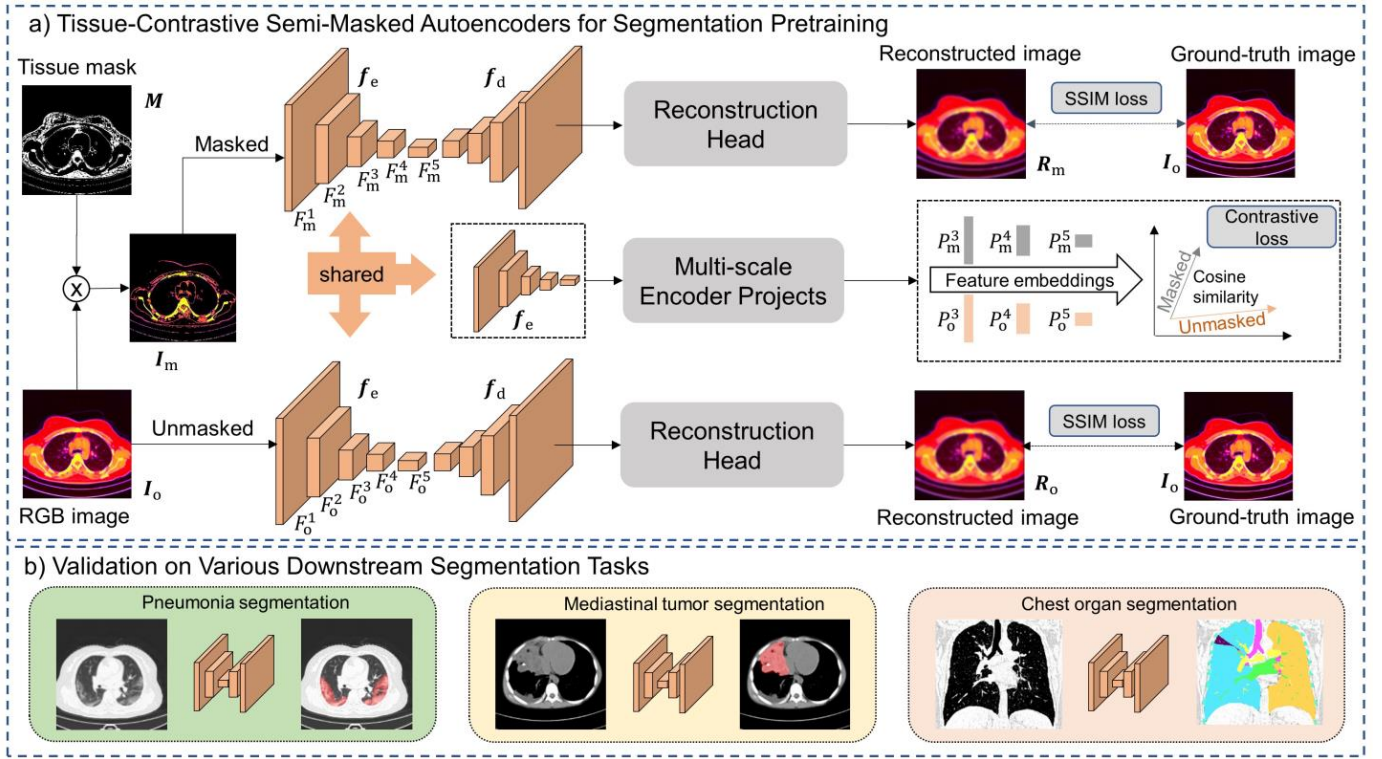

Fig. 2. a) Framework of the proposed TCS-MAE for self-supervised segmentation pretraining in chest CT images. To learn rich anatomical representations, tissue-masked images and their unmasked counterparts are fed into the AE model for tissue-contrastive reconstruction. b) The pretrained models can then be used to enhance the segmentation performance in downstream tasks, such as pneumonia segmentation, mediastinal tumor segmentation and chest organ segmentation.

### C. Hybrid Learning

Recently, several studies have demonstrated that incorporating CL into the AE framework can facilitate more efficient representation learning. For instance, Huang et al.[11] proposed Contrastive Masked Auto-Encoders (CMAE) as stronger vision learners. Tang et al.[23] utilized CL to enhance vision Transformers[24] in learning robust representations from augmented sub-volumes of medical images. Xing et al.[25] combined CL and MIM for 3D medical image segmentation. Although these methods have substantially advanced the field of SSL, they still rely on the patch-based masking strategy and suffer from the upstream-downstream inconsistency issue. Hence, they fall short of being the optimal choice for chest CT pretraining, given the complicated tissue distributions.

## III. METHODS

Figure 2a illustrates the framework of the proposed TCS-MAE. During the training phase, a tissue mask is first generated to mask tissue regions instead of spatial patches in an RGB image that comprises three channels: the lung channel, the mediastinal channel, and the edge channel. Subsequently, both RGB image and its masked counterpart are fed into two AE-based branches for image reconstruction. These two branches share the identical architecture and learnable parameters. Furthermore, to enhance feature learning, a Multi-scale Encoder Projects (MEP) module is designed to project the encoder's feature maps into embeddings, and then a contrastive loss is introduced to the feature embedding pyramids. This loss function mandates consistency between the original RGB image and its masked version within the multiscale feature space, rather than merely ensuring consistency between their reconstructed outputs. Since this dual AE framework features one branch that is masked and another that remains unmasked, we refer to it as a semi-masked AE. Once the model is pretrained, it can be applied to downstream segmentation tasks as demonstrated in Fig. 2b. More details are presented in the following subsections.

### A. Tissue Mask Generation

The tissue mask is obtained by dividing the intensity spectrum of a chest CT image into multiple intervals, following by random masking. For simplicity, we first normalize the HU values into the range of [0, 1]. Then, the dividing operation is performed as follows:

$$S = \{[kW_{\text{intv}}, (k+1)W_{\text{intv}})|k = 0,1,\dots,K-1\}, \quad (1)$$

where $K$ is a hyperparameter used to control the number of intervals. $W_{\text{intv}} = \frac{1.0}{K}$ is the interval width. Finally, a subset of intervals is randomly selected to generate the mask image, which can be formally expressed as follows:

$$M(i,j) = \begin{cases} 0 & if\ I_{\text{norm}}(i,j)\ in\ \forall S_{k \in k_{\text{mask}}} \\ 1 & \text{otherwise} \end{cases}, \quad (2)$$

where $M$ and $I_{\text{norm}}$ denotes the tissue mask and the normalized CT image, respectively. $(i,j)$ is the spatial index of pixels. $k_{mask}$ is a set of interval indexes that are randomly selected to mask. This selection can be expressed as follows:

$$k_{\text{mask}} = \text{RandomChoice}([0,1,\ldots,K-1], \lfloor \rho K \rfloor), \quad (3)$$

where $\rho \in (0,1)$ is a user-defined mask ratio used to control the number of intervals to mask, and $\lfloor * \rfloor$ denotes the floor function.

As previously mentioned, different tissues in CT images are characterized by specific HU ranges. The proposed intensity-based masking strategy can mask entire tissue regions rather than individual patches. This approach can effectively guide the AE model in learning tissue-aware representations, thereby enhancing its ability to recognize different tissue types within CT images.

### B. RGB Image Construction

To concurrently learn both pulmonary and mediastinal tissue representations, the input images for our model are designed with RGB three channels. These channels respectively correspond to the lung image, the mediastinal image and the edge image. The RGB image can be denoted as follows:

$$I_o = (I_{\text{lung}}, I_{\text{medi}}, I_{\text{edge}}), \quad (4)$$

where $I_{\text{lung}}$ and $I_{\text{medi}}$ are the grayscale images that converted from HU values with different clinical window settings using the following function:

$$I_*(i,j) = \frac{I_{ct}(i,j) - l + 0.5w}{w} \times 255, \quad (5)$$

where $I_*$ denotes $I_{\text{lung}}$ or $I_{\text{medi}}$. $I_{ct}(i,j)$ denotes the HU value, $l$ and $w$ are the window level and window width, respectively. The window level and width setting for getting the $I_{\text{lung}}$ is $l = -500$ HU and $w = 1200$ HU, while the setting for the $I_{\text{medi}}$ is $l = 30$ HU and $w = 300$ HU.

Subsequently, the edge image is obtained to highlight the tissue structures. We first convolve the pulmonary images with the horizontal and vertical Sobel kernels[26] as follows:

$$G_x = S_x * I_{\text{lung}}, G_y = S_y * I_{\text{lung}}, \quad (6)$$

where $S_x$ and $S_y$ indicates the Sobel kernels that are defined as:

$$S_x = \begin{bmatrix} -1 & 0 & 1 \\ -2 & 0 & 2 \\ -1 & 0 & 1 \end{bmatrix}, S_y = \begin{bmatrix} -1 & -2 & 1 \\ 0 & 0 & 0 \\ 1 & 2 & 1 \end{bmatrix}. \quad (7)$$

Then, the edge image of lungs is obtained by calculating:

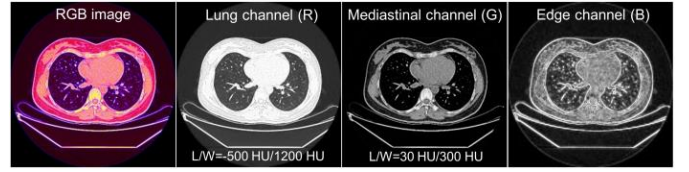

Fig. 3. An example of RGB image composed of three channels: a lung image, a mediastinal image, and an edge image. These images enable the SSL model to learn rich anatomical features in both pulmonary and mediastinal regions.

$$I_{\text{edge}}^{\text{lung}} = \sqrt{G_x^2 + G_y^2}. \quad (8)$$

Similarly, we can obtain the edge image $G_{\text{edge}}^{\text{medi}}$ of the mediastinal image using the above operations. Finally, we get the edge image that presents both pulmonary and mediastinal structural information as follows:

$$I_{\text{edge}}(i,j) = \begin{cases} I_{\text{edge}}^{\text{lung}}(i,j) & if\ I_{\text{edge}}^{\text{lung}}(i,j) > I_{\text{edge}}^{\text{medi}}(i,j) \\ I_{\text{edge}}^{\text{medi}}(i,j) & \text{otherwise} \end{cases}, \quad (9)$$

where $(i,j)$ is the pixel index in the spatial space.

Figure 3 shows an example of RGB image and its three channels. These channels allow the SSL model to discern rich anatomical structures within both pulmonary and mediastinal regions. More importantly, the RGB image is inherently compatible with existing SSL methods that were originally designed for natural images.

### C. Tissue Contrastive Semi-Masked Learning

The minibatch of RGB images and their masked versions are normalized and fed into the AE model for tissue contrastive semi-masked learning. Let the encoder of the AE to be $f_e$ (ResNets[27] or Transformers[28]), multiscale feature maps are generated for each sample:

$$\begin{aligned} F_m^n &= f_e(I_m^n | \theta_e) = [F_m^1, F_m^2, \ldots, F_m^L]^n, \\ F_o^n &= f_e(I_o^n | \theta_e) = [F_o^1, F_o^2, \ldots, F_o^L]^n, \end{aligned} \quad (10)$$

where $I_m^n$ is the $n_{\text{th}}$ masked image in the minibatch that obtained by the element-wise product between $M^n$ and $I_o^n$, i.e., $I_m^n = M^n \otimes I_o^n$. $\theta_e$ denotes the learnable parameters of $f_e$. $L = 5$ represents the number of levels in the feature pyramid, where each level undergoes down sampling with a stride of 2. Then, these feature maps are utilized to reconstruct images, which can be expressed as follows:

$$R_m = \text{Sigmoid}(f_d(F_m | \theta_d)), \quad (11)$$

and

$$R_o = \text{Sigmoid}(f_d(F_o | \theta_d)), \quad (12)$$

where $f_d$ denotes a decoder (e.g., the U-Net decoder) with the model parameters denoted by $\theta_d$. $R_m$ and $R_o$ represents the reconstructed results.

The objective of the reconstruction task is to determine the optimal model parameters $\boldsymbol{\theta}_e^*$ and $\boldsymbol{\theta}_d^*$, which are subject to:

$$<\boldsymbol{\theta}_e^*, \boldsymbol{\theta}_d^*> = \arg\max_{<\boldsymbol{\theta}_e, \boldsymbol{\theta}_d>} \{SSIM(\boldsymbol{R}_m, \boldsymbol{I}_o) + SSIM(\boldsymbol{R}_o, \boldsymbol{I}_o)\}, \quad (13)$$

where $SSIM(*,*)$ is the Structural Similarity Index Measure between two images[29]. This objective can be realized by minimizing the following SSIM loss function:

$$\mathcal{L}_{ssim} = \frac{1}{N}\sum_{n=1}^{N} \begin{Bmatrix} 1 - SSIM(\boldsymbol{R}_m^n, \boldsymbol{I}_o^n) + \\ 1 - SSIM(\boldsymbol{R}_o^n, \boldsymbol{I}_o^n) \end{Bmatrix}. \quad (14)$$

The $SSIM$ between a reconstructed image and its ground-truth image is calculated by:

$$SSIM(\boldsymbol{R}_*^n, \boldsymbol{I}_*^n) = \frac{(2\mu_R\mu_I + \tau_1)(2\delta_{RI} + \tau_2)}{(\mu_R^2 + \mu_I^2 + \tau_1)(\delta_R^2 + \delta_I^2 + \tau_2)}, \quad (15)$$

where $\mu_*$ and $\delta_*$ are the mean and variance of the images $\boldsymbol{R}_*^n$ and $\boldsymbol{I}_*^n$ (the subscript dot indicates pulmonary or mediastinal images). $\delta_{RI}$ is the covariance between $\boldsymbol{R}_*^n$ and $\boldsymbol{I}_*^n$. $\tau_1$ and $\tau_2$ are predefined constant values.

Meanwhile, the MEP module projects the encoder's feature maps of each sample into feature embeddings. This operation can be formally described as follows:

$$\begin{aligned} \boldsymbol{P}_m^n &= \left[f_p^1(F_m^1), f_p^2(F_m^2), \ldots, f_p^L(F_m^L)\right]^n, \\ \boldsymbol{P}_o^n &= \left[f_p^1(F_o^1), f_p^2(F_o^2), \ldots, f_p^L(F_o^L)\right]^n, \end{aligned} \quad (16)$$

where $f_p^l$ denotes the $l_{th}$ projection layer which consists of a flatten layer followed by a fully connected layer. $f_p^l(F_*^l) \in R^{1\times D_l}$ indicates an embedding, where $D_l = 128$ in this study. The extracted embeddings are then used to calculating the contrastive loss. We follow the SimCLR[14] to define the contrastive loss as

$$\mathcal{L}_{con} = -\frac{1}{N}\sum_{n=1}^{N}\sum_{l=1}^{L}\log\frac{\exp\left(sim(P_m^{n,l}, P_o^{n,l})/\gamma\right)}{\sum_{k=1}^{2N}\mathbb{1}_{k\neq n}\exp\left(sim(P_m^{n,l}, P_o^{k,l})/\gamma\right)}, \quad (17)$$

where $\gamma$ is a learnable temperature parameter. $\mathbb{1}_{k\neq n}$ is an indicator function. $P_*^{n,l} = f_p^l(F_*^l)^n$ indicates the $l_{th}$ layer embedding of the $n_{th}$ sample in the minibatch. $sim(u,v) = u^T v/\|u\|\|v\|$ denotes the cosine similarity between vectors $u$ and $v$. Finally, the overall learning target is a weighted sum of the reconstruction loss and the contrastive loss:

$$\mathcal{L}_{total} = \mathcal{L}_{ssim} + \lambda\mathcal{L}_{con}, \quad (18)$$

where $\lambda$ is a hyperparameter used to adjust the contribution of the consistence term. $\lambda$ is default set to 1.0 in this study.

### D. Downstream Segmentation Applications

To validate the efficacy of our pretraining approach, we conduct finetuning of the pretrained models for a variety of downstream segmentation tasks, including the delineation of pneumonia, mediastinal tumors, and thoracic organs as depicted in Fig. 2b. To accommodate the downstream segmentation tasks with ease, the pretrained TCS-MAE models are founded on the U-Net architecture. Nevertheless, the pretrained model's output layer contains only a single channel activated by a Sigmoid function. We modify the output layer to multiple channels with SoftMax activation, when adapting it to a multiclass segmentation task. During the training phase of the downstream models, their initial parameter values are inherited from the pretrained models. Then, a modest learning rate is employed to finetune the models over a restricted number of iterations.

## IV. EXPERIMENTS

### A. Datasets

Four datasets were built for the development and validation of our method (see Table I), including:

1) *ChestCT-SSL*: This is a private large-scale dataset for self-supervised pretraining. It includes a comprehensive collection of 6027 normal chest CT scans, amounting to 635,120 slices. These scans were acquired from multiple cooperative hospitals and are characterized by the absence of reported lesions. The slice thickness varies from 1.5mm to 5mm, which covers a wide spectrum of reconstruction protocols typically encountered in clinical practice.

2) *Pneu-Seg*: This private dataset consists of 501 densely annotated chest CT scans collected from pneumonia cases caused by four distinct pathogen types, including COVID-19, fungal infection, gram-negative bacterial infection, and gram-positive bacterial infection.

3) *Medi-Seg*: This is a private dataset with 742 chest CT scans for mediastinal lesion segmentation. It contains

TABLE I
THE DATASETS FOR VALIDATING THE PROPOSED METHOD

| Dataset | Samples | Train/Test | Annotated lesions/tissues | Task |
|---|---|---|---|---|
| ChestCT-SSL | 6027 | 6027/0 | Normal scans without lesions | Segmentation pretraining |
| Pneu-Seg | 501 | Five-fold cross validation | COVID-19, fungal, and bacterial infections | Pneumonia segmentation |
| Medi-Seg | 742 | Five-fold cross validation | Neuroma, teratoma, thymoma, Germinoma, and lymphadenoma | Mediastinal tumor segmentation |
| Total-Segmentator[30] | 538 | Five-fold cross validation | Pulmonary artery, esophagus, trachea, left lung upper lobe, left lung lower lobe, right lung upper lobe, right lung middle lobe, right lung lower lobe, heart myocardium, left heart atrium, left heart ventricle, right heart atrium, right heart ventricle, vertebrae, rib, and scapula | Chest organ segmentation |

annotations in the form of masks for lesions originating from five common seen mediastinal tumor types: neuroma, teratoma, thymoma, germinoma, and lymphadenoma.

4) *Total-Segmentator*[30]: This is a public large-scale dataset comprises 1128 CT scans with masks annotated for 104 classes of anatomical tissues. In this study, we select 538 scans with the chest body parts for validating our method, which includes pulmonary artery, esophagus, trachea, left lung upper lobe, left lung lower lobe, right lung upper lobe, right lung middle lobe, right lung lower lobe, heat myocardium, left heart atrium, left heart ventricle, right heart atrium, right heart ventricle, vertebrae, rib, and scapula (16 categories).

### B. Implementation Details

1) *Pretraining*: The U-Net architecture is implemented as the backbone of the AE model, with SegFormers[28] serving as the encoder. The encoder outputs five layers of feature maps for each RGB image and its masked version: $[F_o^1, F_o^2, F_o^3, F_o^4, F_o^5]$ and $[F_m^1, F_m^2, F_m^3, F_m^4, F_m^5]$. By default, the last two pairs of feature maps (i.e., $[F_o^4, F_o^5]$ and $[F_m^4, F_m^5]$) are utilized for the contrastive learning in the MEP module. We leverage Pytorch to train these models on an NVIDIA A800 GPU equipped with 80GB of memory. The multi-task loss (see Eq.18) is minimized using the Adam optimizer[31] over 60 epochs, with an initial learning rate of 0.0001 that decays exponentially at a rate of 0.96 per epoch. The batch size is set to 30, and the input images are scaled to a resolution of $256 \times 256$ pixels.

2) *Finetuning*: For the tasks involving pneumonia and mediastinal tumors, the models are finetuned for binary segmentation, i.e., differentiating lesion regions from non-lesion background. Consequently, no alterations to the model architecture are necessitated. In the case of organ segmentation task, the output layer is revised to comprise 17 channels, which employs the Softmax activation function to facilitate multiclass segmentation (background plus 16 organ classes). The Cross Entropy combined with Dice loss is optimized using an Adam optimizer during the finetuning process. A reduced learning rate of 0.000001 is adopted, and the number of epochs is set to 50.

### C. Evaluation Metrics

The downstream segmentation performance is validated using two metrics: the Dice Similarity Coefficient (*DSC*) and the Hausdorff Distance (*HD*). The *DSC* score quantifies the overlap between the predicted masks and their ground-truth masks, whereas the *HD* score measures the largest minimum distance between two contour sets. Superior performance is indicated by a higher *DSC* score and a lower *HD* score. To ensure the reliability of the results, we perform five-fold cross validation, and the overall performance is assessed by calculating the mean and standard deviation of the metrics across five folds.

## V. RESULTS AND DISCUSSION

### A. Comparison with SOTA SSL Methods

In this study, we benchmark our method against a spectrum of SSL approaches, including both contrastive and generative

TABLE II
FIVE-FOLD BINARY SEGMENTATION RESULTS OF PNEUMONIA AND MEDIASTINAL TUMORS

| Method | *Pneu-Seg* DSC (%) | HD | *Medi-Seg* DSC (%) | HD |
|---|---|---|---|---|
| Baseline | 66.23±0.49 | 5.44±0.16 | 62.42±1.90 | 3.39±0.13 |
| SimCLR[14] | 67.72±0.57 | 5.40±0.05 | 64.07±1.60 | 3.48±.013 |
| DINO[15] | 66.74±0.69 | 5.36±0.05 | 63.62±2.01 | 3.46±0.09 |
| MoCo[16] | 66.63±0.76 | 5.44±0.15 | 63.20±1.59 | 3.37±0.11 |
| AE[32] | 68.78±0.60 | 5.40±0.05 | 65.19±1.62 | 3.42±0.12 |
| DAE[33] | 69.95±0.93 | 5.30±0.06 | 67.81±0.84 | 3.37±0.12 |
| MAE[9] | 70.25±0.91 | 5.46±0.05 | 68.10±2.16 | 3.29±0.12 |
| CMAE[11]† | 70.87±0.97 | 5.31±0.06 | 68.88±1.39 | 3.26±0.11 |
| CAE[34] | 69.18±0.84 | 5.48±0.05 | 65.20±2.40 | 3.47±0.09 |
| MaskFeat[35] | 66.64±0.86 | 5.80±0.05 | 62.59±1.44 | 3.63±0.08 |
| CLGL[19] | 67.96±0.63 | 5.39±0.05 | 64.85±1.71 | 3.34±0.10 |
| TransVW [36] | 67.21±0.66 | 5.41±0.08 | 65.52±1.13 | 3.55±0.13 |
| HybMIM[25]† | 69.96±0.73 | 5.35±0.06 | 67.35±1.65 | 3.32±0.12 |
| TCS-MAE† | **71.04±0.47** | **5.28±0.05** | **72.78±1.16** | **3.21±0.12** |

The best score in each column is shown in bold. Baseline: The U-Net trained from scratch. Mit-b2[28] is used as the encoder for the U-Net. †indicates the MIM with contrastive learning. Our TCS-MAE: $K = 8, \rho = 75\%$; MEP for Contrastive learning: $[F_o^4, F_o^5]$ and $[F_m^4, F_m^5]$.

methodologies that were initially developed for modeling natural images. The former includes SimCLR[14], DINO[15], and MoCo[16], while the latter encompasses AE[32], DAE[33], MAE[9], CMAE[11], CAE[34], and MaskFeat[35].

We also compare our method with three SSL methods that were specifically developed for medical image modeling: CLGL[19], TransVW[36], and HybMIM[25]. The TransVM learns rich anatomical patterns by restoring and classifying masked visual words. These visual words are autonomously discovered via an anatomical consistency learning procedure. Likewise, the HybMIM also capitalizes on restoration and classification tasks to learn anatomical representations from image patches, but it enhances the feature learning through the incorporation of a dropout-based contrastive learning strategy.

1) *Lesion segmentation*: Table II presents the segmentation results of pneumonia and mediastinal tumors. From the table, three notable observations can be made as follows: First, All SSL methods can improve the performance of both downstream segmentation tasks. Notably, generative approaches outperform contrastive methods because they pretrain both encoder and decoder for dense predictions. For example, the baseline model (i.e., the U-Net trained from scratch) only achieves a *DSC* score of 66.23±0.49% for pneumonia segmentation. The contrastive SimCLR[14] slightly improves this to 67.72±0.57%, whereas the generative MAE[9] significantly boosts the score to 70.25 ±0.91%.

Second, integrating MIM with contrastive learning indeed enhances feature representations. For instance, the CMAE[11] surpasses the MAE[9], and the HybMIM[25] outperforms the TransVW[36]. However, the HybMIM is even inferior to the MAE. This phenomenon reveals that the MIM strategy remains pivotal in feature learning. To alleviate the issue of high computational demand in medical images, the HybMIM

7TABLE III
FIVE-FOLD MULTICLASS SEGMENTATION RESULTS OF CHEST ORGANS ON THE *TOTAL-SEGMENTATOR* DATASET

| Method | Tubular tissues DSC (%) | HD | Left lung DSC (%) | HD | Right lung DSC (%) | HD | Heart DSC (%) | HD | Bones DSC (%) | HD |
|---|---|---|---|---|---|---|---|---|---|---|
| Baseline (scratch) | 84.60±6.85 | 2.66±1.12 | 95.74±3.42 | 2.36±0.86 | 95.67±3.87 | 2.31±0.92 | 92.27±5.01 | 3.41±0.89 | 82.34±5.20 | 3.08±1.02 |
| SimCLR[14] | 84.83±6.84 | 2.62±1.15 | 95.74±3.35 | 2.42±0.88 | 95.64±3.81 | 2.35±0.94 | 92.09±4.91 | 3.63±0.85 | 81.18±5.66 | 3.17±0.98 |
| DINO[15] | 84.76±6.90 | 2.56±1.12 | 95.69±3.40 | 2.51±0.91 | 95.61±3.85 | 2.46±0.95 | 91.93±4.79 | 3.69±0.83 | 80.80±5.65 | 3.25±0.98 |
| MoCo[16] | 84.47±6.85 | 2.70±1.15 | 95.73±3.40 | 2.38±0.92 | 95.66±3.84 | 2.30±0.92 | 92.15±5.16 | 3.47±0.88 | 81.85±5.34 | 3.13±1.01 |
| AE[32] | 85.09±7.12 | 2.59±1.15 | 95.74±3.39 | 2.34±0.86 | 95.65±3.91 | 2.28±0.93 | 92.09±4.94 | 3.51±0.87 | 81.78±5.53 | 3.09±1.04 |
| DAE[33] | 85.30±6.66 | 2.65±1.17 | 95.74±3.38 | 2.33±0.87 | 95.71±3.78 | 2.24±0.91 | 92.21±4.83 | 3.47±0.88 | 82.69±5.19 | 3.00±1.05 |
| MAE[9] | 85.40±6.35 | 2.67±1.13 | 95.75±3.30 | 2.24±0.81 | 95.77±3.72 | 2.15±0.88 | **92.48±5.02** | **3.23±0.91** | 83.35±5.02 | 2.91±1.08 |
| CMAE[11] | 85.02±7.06 | 2.57±1.13 | 95.76±3.34 | 2.34±0.85 | 95.74±3.79 | 2.25±0.90 | 92.11±4.85 | 3.51±0.87 | 82.47±5.34 | 2.99±1.05 |
| CAE[34] | 85.31±6.93 | **2.54±1.16** | 95.77±3.36 | 2.37±0.87 | 95.70±3.74 | 2.32±0.93 | 92.00±4.92 | 3.54±0.85 | 81.68±5.52 | 3.13±1.03 |
| MaskFeat[35] | 84.48±6.75 | 2.68±1.14 | 95.77±3.33 | 2.54±0.87 | 95.69±3.79 | 2.30±0.92 | 92.24±4.90 | 3.43±0.89 | 81.70±5.39 | 3.16±1.02 |
| CLGL[19] | 84.54±6.84 | 2.70±1.14 | 95.77±3.27 | 2.36±0.86 | 95.69±3.83 | 2.30±0.90 | 92.25±5.05 | 3.40±0.88 | 81.87±5.43 | 3.12±1.01 |
| TransVW[36] | 85.30±6.43 | 2.77±1.13 | 94.70±3.20 | 2.34±0.83 | 94.73±3.73 | 2.23±0.89 | 92.08±5.12 | 3.33±0.81 | 83.26±5.12 | 2.89±1.04 |
| HybMIM[25] | 84.68±6.65 | 2.87±1.14 | 94.64±3.33 | 2.36±0.87 | 94.57±3.79 | 2.35±0.91 | 91.58±5.22 | 3.36±0.78 | 83.15±5.31 | 2.90±1.06 |
| Our TCS-MAE | **86.02±6.13** | 2.62±1.13 | **95.79±3.18** | **2.22±0.82** | **95.80±3.59** | **2.10±0.87** | 92.28±4.84 | 3.29±0.89 | **84.08±4.73** | **2.82±1.12** |

The best score in each column is shown in bold. Baseline is the U-Net trained from scratch. Mit-b2[28] is used as the encoder for the U-Net. Our TCS-MAE: $K = 8, \rho = 75\%$; MEP for Contrastive learning: $[F_o^4, F_o^5]$ and $[F_m^4, F_m^5]$. **Tubular tissues**: pulmonary artery, esophagus, trachea; **Left lung**: left lung upper lobe, left lung lower lobe; **Right lung**: right lung upper lobe, right lung middle lobe, right lung lower lobe; **Heart**: left heart atrium, left heart ventricle, right heart atrium, right heart ventricle; **Bones**: vertebrae, rib, and scapula.

employs a dynamic partial region selection strategy to reconstruct only partial image regions. This solution can accelerate the training process but may compromise the effectiveness of feature learning.

Third, our TCS-MAE ranks first in boosting the segmentation performance, particularly for the mediastinal tumor segmentation task. Compared to the lungs, mediastinal regions typically present more complex anatomical structures. Consequently, segmenting mediastinal tumors is much harder than pneumonia segmentation. Promisingly, the *DSC* scores for tumor segmentation achieved by existing methods are much lower than those for pneumonia segmentation. Surprisingly, our method, on the contrary, achieves better results in mediastinal tumor segmentation. The *DSC* score for tumor segmentation is 72.78 ± 1.16%, but the counterpart for pneumonia segmentation is 71.04±0.47%. We believe this is because our tissue-based masking strategy guides the model to learn more robust tissue-aware representations. Therefore, its advantages are more evident in the task of mediastinal tumor segmentation.

2）*Organ segmentation*: To further validate our method, we conduct experiments on chest organ segmentation using the *Total-Segmentator* dataset[30]. Because there are too many tissue types (a total of 16 categories), we group the tissues into five categories and only calculate metrics on these groups: 1) Tubular tissues (pulmonary artery, esophagus, and trachea), 2) left lung (left lung upper lobe, left lung lower lobe), 3) right lung (right lung upper lobe, right lung middle lobe, right lung lower lobe), 4) heart (left heart atrium, left heart ventricle, right heart atrium, right heart ventricle), and 5) bones (vertebrae, rib, and scapula). The results are presented in Table III.

From table III, a similar conclusion can be drawn that most SSL methods enhance the segmentation performance. However, the improvements are not as significant as those observed in the lesion segmentation tasks (see Table II). We attribute this phenomenon to the fact that, unlike lesions with substantial variation, organs tend to display consistent characteristics across different individuals. This consistency enables the baseline model to achieve commendable results even when trained on a set of 538 meticulously annotated samples. Compared to existing SSL methods, our method still achieves better performance over most tissues, except for the heart. The *DSC* score achieved by our method for heart is 92.28±4.84%, slightly lower than MAE's 92.48±5.02%.

3）*Performance analysis*: Figure 4 plots the online evaluation of *DSC* scores corresponding to the three segmentation tasks, which demonstrates that both CMAE and our TCS-MAE can significantly aid the improvement of segmentation metrics. However, compared to the CMAE, our method exhibits more distinct advantages, particularly in the first 10 epochs. Besides, Fig. 4 reveals that the baseline model also attains high segmentation accuracy for organ segmentation. Notably, after just the first epoch, the baseline model achieves a *DSC* score exceeding 80%. This phenomenon further confirms the preceding conclusion: the consistency in tissue characteristics across individuals facilitates organ segmentation, consequently diminishing the prominence of SSL in this task.

It is hypothesized that a proficient SSL learner for chest CT images would possess a strong ability to distinguish tissues well. Only through such capability can it more effectively recover the masked images. To analysis why our method outperforms other methods, we calculate the *SSIM* scores between images and their reconstructions from pretrained models of CMAE and TCS-MAE. Fig. 5 shows the line plots using 1000 samples. Evidently, our TCS-MAE achieves much higher *SSIM* scores compared to CMAE. Even without masking the input images for the CMAE (see CMAE* in Fig. 5), our TCS-MAE, utilizing masked inputs, can achieve comparable *SSIM* levels. These findings underscore the powerful tissue representation ability of our pretrained model.

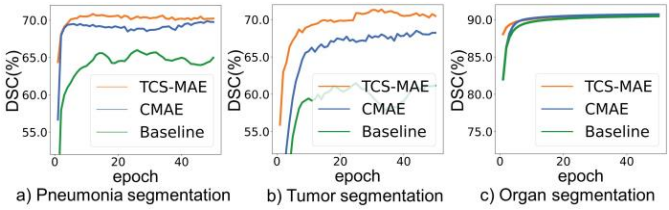

Fig. 4. The online evaluation of *DSC* scores for segmenting a) pneumonia, b) mediastinal tumors, and c) organs. Our TCS-MAE can significantly accelerate the improvement of *DSC* scores. Benefiting from the consistency of tissue characteristic across individuals, the baseline model can also attain high segmentation score quickly for organ segmentation as demonstrated in c).

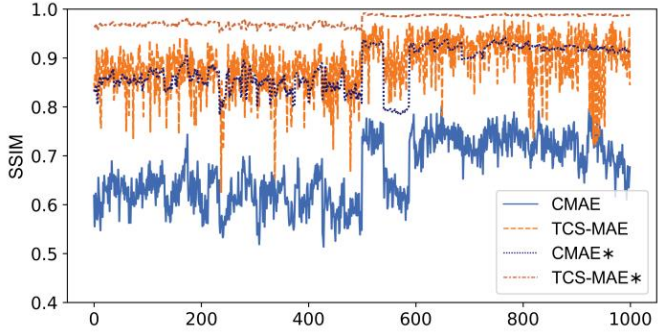

Fig. 5. Line plots of *SSIM* scores calculated between images and their reconstructions from pretrained CMAE and our TCS-MAE models. A random selection of one thousand samples is made to generate these plots. The asterisk indicates the inputs for the pretrained models are unmasked images. Evidently, our TCS-MAE, benefiting from its powerful tissue representation ability, achieves much higher *SSIM* scores for both masked and unmasked inputs.

Finally, we visualize the upstream reconstruction results on abnormal samples and their corresponding downstream segmentation results. Fig. 6 presents three representative cases for the CMAE and our TCS-MAE, which intuitively reveals that our method more accurately restores masked images than does the CMAE. This advantage stems from our tissue-masking strategy, which enables the model to learn more robust tissue-aware representations compared to the patch-based masking strategy. Such capability enhances the efficiency of downstream segmentation tasks.

### B. Ablation Study

1) *Masking hyperparameters*: Two hyperparameters, i.e., the number of HU intervals $K$ (see Eq. 1) and the masking ratio $\rho$ (see Eq. 3), are defined to regulate the masking degree of tissues, which may affect the efficacy of feature learning and thus influence the performance of downstream segmentation. To assess the impact, ablation experiments are performed using various configurations with $K$ values of 4, 6, 8, and $\rho$ values set at 15%, 35%, 55%, and 75%. Fig. 7 shows the box plots for five-fold cross-validation in pneumonia and mediastinal tumor segmentation tasks. It can be observed that $K = 8$ is a better option for both segmentation tasks. Besides, for the pneumonia segmentation task, the optimal setting is $\rho = 35\%$, whereas for mediastinal tumor segmentation, the best results are achieved with $\rho = 75\%$. Compared to the lung regions, which are primarily composed of air, mediastinal areas contain more complex anatomical structures with a broader range of HU. It is supposed that a higher masking ratio $\rho$ enables the model to learn more fine-grained tissue features from the mediastinum.

2) *Masking strategy*: To validate the benefits of our tissue-masking strategy, we carry out ablation studies with two setups: "Patch+Unmask" and "Tissue+Unmask". Here, 'Patch' and 'Tissue' refer to the MAE branch that utilizes the patch-based and tissue-based masking, respectively. The term 'Unmask' pertains to the AE branch. The results for pneumonia and mediastinal tumor segmentation tasks are tabulated in Table IV. It is evident that substituting tissue masking with patch masking in the MAE branch of our framework results in a decline in performance of the tumor segmentation task, where there is a decrease in the *DSC* value by 1.09%. However, in the case of pneumonia segmentation, the "Patch+Unmask" configuration attains a *DSC* score of $71.05 \pm 0.92\%$, which is marginally

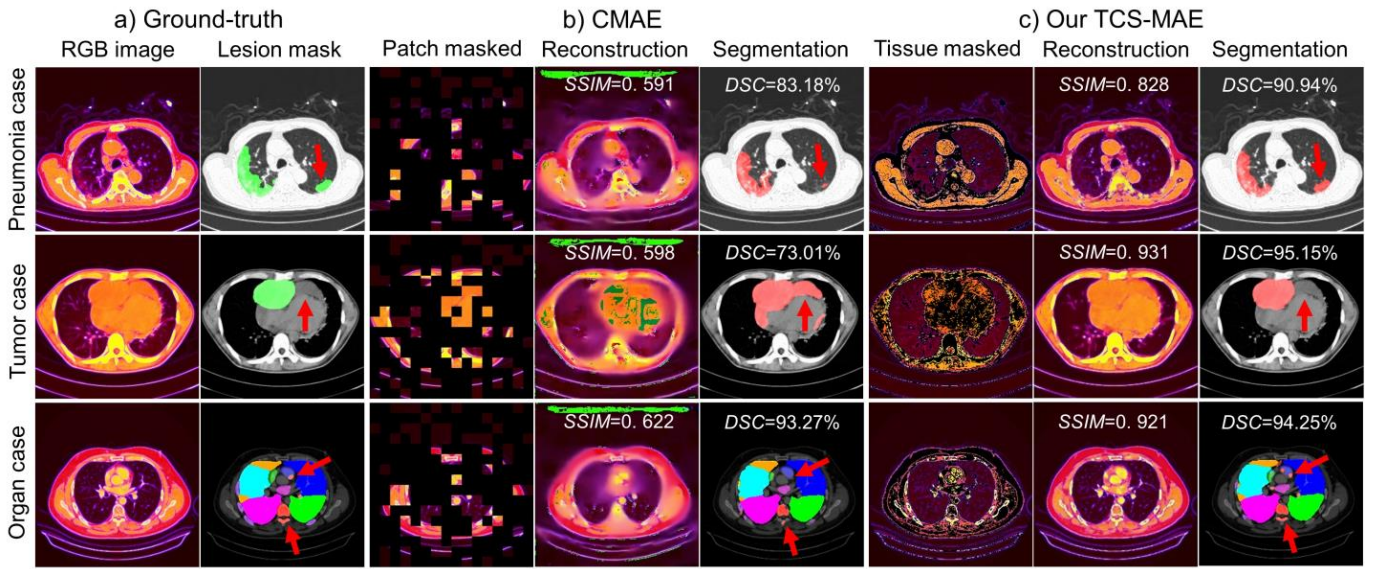

Fig. 6. Three visualization cases correspond to pneumonia, mediastinal tumor, and organ. Compared to the patch-masking strategy that used by CMAE, Our TCS-MAE can learn better tissue-aware representations for reconstruction using the tissue-masking strategy, which facilitates the downstream segmentation tasks. The red arrows indicate the regions where the CMAE has performed incorrect segmentation.

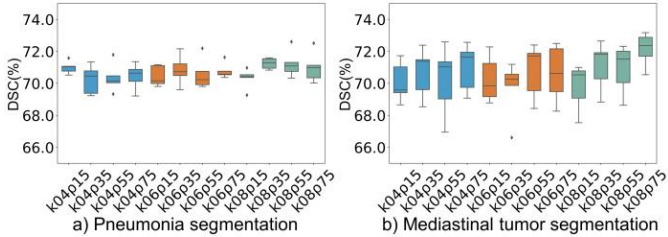

Fig. 7. Different settings of masking hyperparameters result in varying performance. For pneumonia segmentation, the optimal setting is $K = 8$ and $\rho = 35\%$, whereas for mediastinal tumor segmentation, the best results are achieved with $K = 8$ and $\rho = 75\%$.

TABLE IV
IMPACT OF MASKING STRATEGY ON SEGMENTATION PERFORMANCE OF PNEUMONIA AND MEDIASTINAL TUMORS

|  | Pneu-Seg |  | Medi-Seg |  |
| --- | --- | --- | --- | --- |
| Masking strategy | DSC (%) | HD | DSC (%) | HD |
| Patch+Unmask | **71.05±0.92** | **5.19±0.07** | 71.37±0.92 | 3.26±0.12 |
| Tissue+Unmask | 71.04±0.47 | 5.28±0.05 | **72.78±1.16** | **3.21±0.12** |

The best score in each column is shown in bold. Patch masking: masking size is 16 and masking ratio is 75%; Tissue masking: masking size is 8 (i.e., $K = 8$) and masking ratio is 75% (i.e., $\rho = 75\%$).

TABLE V
IMPACT OF EMBEDDING SCALES ON SEGMENTATION PERFORMANCE OF PNEUMONIA AND MEDIASTINAL TUMORS

|  | Pneu-Seg |  | Medi-Seg |  |
| --- | --- | --- | --- | --- |
| MEP with embedding scales for CL | DSC (%) | HD | DSC (%) | HD |
| Scales 0 | **71.72±0.52** | 5.23±0.05 | 72.29±1.52 | 3.22±0.12 |
| Scales 1 | 71.68±0.44 | 5.25±0.05 | 72.46±0.94 | 3.24±0.09 |
| Scales 2 | 71.04±0.47 | 5.28±0.05 | **72.78±1.16** | 3.21±0.12 |
| Scales 3 | 68.21±0.39 | 5.46±0.05 | 72.43±1.35 | **3.21±0.11** |

The best score in each column is shown in bold. MEP: Multi-scale Encoder Projects module; CL: Contrastive Learning.

superior to the 71.04±0.47% achieved by the "Tissue+Unmask" configuration. The above results further confirm the superiority of tissue masking for modeling anatomical structures in the mediastinal regions of chest CT images.

*3) Contrastive scales*: All the above experiments are based on two scales of feature embeddings for contrastive learning (CL) in the MEP module. To assess the impact of multiscale feature embeddings, we also pretrained several TCS-MAE models with different embedding scales: Scales 0 without CL, Scales 1 with $<[F_o^5], [F_m^5]>$, Scales 2 with $<[F_o^4, F_o^5], [F_m^4, F_m^5]>$, and Scales 3 with $<[F_o^3, F_o^4, F_o^5], [F_m^3, F_m^4, F_m^5]>$. The downstream segmentation metrics are tabulated in Table V. A trend can be observed that performance becomes worse, especially in pneumonia segmentation, when increasing the number of embedding scales for CL. This could be attributed to the model's encoder being increasingly biased towards learning contrastive features, thus reducing the decoder's reconstruction capability. Fig. 8 demonstrates that the convergence of reconstruction loss becomes slower, as the number of contrastive scales increases. However, for the mediastinal segmentation, utilizing Scales 2 yields the highest *DSC* score. Considering that mediastinal regions contain rich anatomic structures of diverse sizes and shapes, the setting of Scales 2 can aid model in learning more fined-grained tissue information within these areas, even though it results in slower convergence.

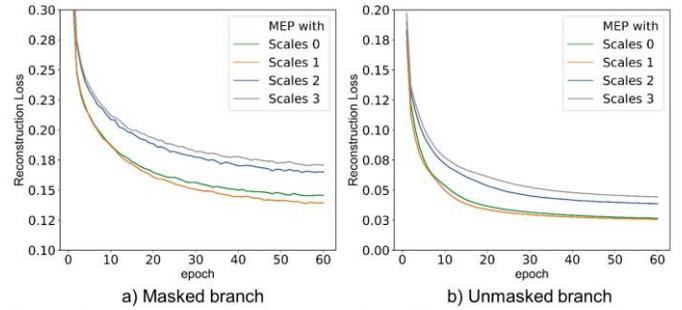

Fig. 8. Reconstruction loss curves of our TCS-MAE corresponding to different settings of embedding scales for CL in the MEP module. More than one scales leads to slower convergence.

### C. Limitations and Future Work

The above experiments substantiate that our TCS-MAE approach can indeed enhance deep learning models in learning tissue representations from unlabeled chest CT images. However, there are some shortcomings in current study. Firstly, to conveniently validate the effectiveness of our method and compare it with existing SSL methods, we have implemented TCS-MAE in a 2D version. This implementation may overlook the contextual 3D tissue information present in CT volumes. Secondly, as revealed by the experimental results, our method excels at learning complex tissue regions such as mediastinum in CT images. We believe that it can also be extend beyond just chest CT scans to various other medical imaging of different body parts. However, the current study only focuses on chest CT. In the future, we will implement a 3D variant of TCS-MAE and verify it on a broader range of medical images across diverse tasks.

## VI. CONCLUSION

This paper introduces a novel self-supervised learning method named Tissue-Contrastive Semi-Masked Autoencoder (TCS-MAE) which aims to improve the representation quality of CT images. In TCS-MAE, we design a tissue-based masking strategy and a semi-masked dual-AE framework to address the shortcomings of existing MIM-based methods for modeling CT images. Comprehensive experiments on multiple segmentation tasks demonstrate that, through these designs, our method can significantly boost the AE model in capturing tissue-aware features, thus achieving best performance over exiting SSL.

### ACKNOWLEDGMENT

This work is supported by the Supercomputing Center of Hangzhou City University.### REFERENCES

[1] J. Wang, Y. Bao, Y. Wen, H. Lu, H. Luo, Y. Xiang, X. Li, C. Liu, and D. Qian, "Prior-Attention Residual Learning for More Discriminative COVID-19 Screening in CT Images," *IEEE Trans. Med. Imaging*, vol. 39, no. 8, pp. 2572–2583, 2020.

[2] S. A. Harmon, T. H. Sanford, S. Xu, E. B. Turkbey, H. Roth, Z. Xu, D. Yang, A. Myronenko, V. Anderson, A. Amalou, M. Blain, M. Kassin, D. Long, N. Varble, S. M. Walker, U. Bagci, A. M. Ierardi,